\documentclass[pre,aps,preprint,showpacs]{revtex4}
\usepackage{revsymb,amsmath}
\usepackage{graphicx}
\newcommand{\bel}[1]{\begin{equation}\label{#1}}

\newcommand{\exval}[1]{\mbox{$\langle \, {#1}\, \rangle$}}
\def\bbbz{{\mathchoice {\hbox{$\sf\textstyle Z\kern-0.4em Z$}}
{\hbox{$\sf\textstyle Z\kern-0.4em Z$}}
{\hbox{$\sf\scriptstyle Z\kern-0.3em Z$}}
{\hbox{$\sf\scriptscriptstyle Z\kern-0.2em Z$}}}}
\def\be{\begin{equation}}
\def\ee{\end{equation}}
\def\bea{\begin{eqnarray}}
\def\eea{\end{eqnarray}}
\def\ba{\begin{array}}
\def\ea{\end{array}}
\begin{document}
\author{Gunter M. Sch\"utz}
\affiliation{Institut f\"ur Festk\"orperforschung, Forschungszentrum J\"ulich,
D-52425 J\"ulich, Germany}
\email{g.schuetz@fz-juelich.de}
\author{Steffen Trimper}
\affiliation{Fachbereich Physik, Martin-Luther-Universit\"at, D-06099 Halle,
Germany}
\email{trimper@physik.uni-halle.de}
\title{Elephants can always remember: Exact long-range memory effects
in a non-Markovian random walk}
%\draft
\date{\today }

\begin{abstract}
We consider a discrete-time random walk where the random increment at time 
step $t$ depends on the full history of the process. We calculate exactly the mean and
variance of the position and discuss its dependence on the initial condition
and on the memory parameter $p$. At a critical value $p_c^{(1)}=1/2$ where
memory effects vanish there is a transition from a weakly localized regime 
(where the walker returns to its starting point) to an escape regime.
Inside the escape regime there is a second critical value where the random walk becomes superdiffusive.
The probability distribution is shown to be governed by a non-Markovian
Fokker-Planck equation with hopping rates that depend both on time and on the 
starting position of the walk. On large scales the memory organizes itself 
into an effective harmonic oscillator potential for the random walker with a
time-dependent spring constant $k = (2p-1)/t$. The solution of this
problem is a Gaussian distribution with time-dependent mean
and variance which both depend on the initiation of the process.
\end{abstract}
\maketitle

Memory effects in non-Markovian stochastic processes are often
incorporated heuristically on a coarse-grained scale into time evolution
equations for physical observables, recently discussed in \cite{tsi01,ma02,fr00}. 
A fundamental concept of a non-Markovian process is the continuous-time random walk introduced 
already some years ago \cite{mw}. This theory has numerous important applications, studied for 
instance in \cite{ah} or recently in \cite{im}. A broad variety of examples in biology is 
analyzed in \cite{mur}. Alternatively, one may derive 
a formally exact evolution equation for an observable by a projection 
mechanism \cite{Zwan61,Mori73} and then apply some approximation scheme for 
the solution of the equation. Techniques of this type are used 
in the context of diffusive dynamics where memory effects may lead to 
anomalous diffusion or even localization. In recent studies this has been 
demonstrated within an one-loop renormalization group approach \cite{ss}, 
by other analytical studies \cite{st1} and confirmed by numerical methods 
\cite{bst}. A more fundamental approach to anomalous diffusion based on a 
nonequilibrium statistical description is already discussed in \cite{wt}.   
While very successful both phenomenologically and in predicting 
interesting new memory effects \cite{tr}, two intrinsic shortcomings of 
these traditional approaches deserve attention: There is no quantitative 
control over the error induced by approximations and the microscopic origin 
of the memory term is frequently obscured. In particular, there is usually
no simple transition from an explicit and physically motivated
non-Markovian noise term to an associated non-Markovian evolution equation
for the probability density which is often employed in the framework
described above \cite{Hang77}. It is the aim of this paper to
skirt these problems in the investigation of memory effects in diffusive
motion by first defining a simple ``microscopic'' non-Markovian stochastic 
dynamics for the random walk on lattice scale, then passing to 
an evolution equation for the probability distribution (in formal
analogy to passing to a Fokker-Planck equation (FPE) in the case of Markov 
processes) and finally coarse graining by taking the limit of large space- 
and time scales. As the result we get a FPE with a time-dependent 
drift term. Recently stochastic processes leading to FPE with time-dependent coefficients, 
are discussed by several authors \cite{lm,dh,mm}. In view of those more heuristic 
approaches our model yields a more microscopic foundation for a special realization of a 
FPE with a time-dependent term.\\  
Specifically we are interested in conditions under which an unbounded memory
can induce qualitative changes in the distribution of the position 
as compared to the Markovian case with Gaussian distribution on large
space and time scales. It is well-known from the self-avoiding 
walk (which is a rare example for a random walk with unbounded memory
where detailed exact results are known \cite{SAW1,SAW2})
that the memory of the previously visited sites changes the scaling 
behavior of the distribution and leads to a superdiffusive
mean square displacement. Here we investigate how a very different
unbounded memory affects the random walk statistics and {\it induces}
a transition to superdiffusive behavior. 
For definiteness and simplicity of notation we mainly consider a 
one-dimensional random walk $X_t \in \bbbz$ on the infinite lattice. The 
random walk starts at some specific point $X_0$ at time $t_0=0$
and has a complete memory of its whole history. In allusion to the
traditional saying that elephants can always remember we shall refer to the 
random walker as elephant. In each discrete time step the elephant
moves one step to the right or left respectively (simple random walk), so 
the stochastic evolution equation is given by
\bel{1}
X_{t+1} = X_t + \sigma_{t+1}
\ee
where $\sigma_{t+1} = \pm 1$ is a random variable. The memory consists of
the set of random variables $\sigma_{t'}$ at previous time steps
which the elephant remembers as follows:\\
{\it (D1)} At time $t+1$ a number $t'$ from the set $\{1,2,\dots,t\}$
is chosen randomly with uniform probability $1/t$.\\
{\it (D2)} $\sigma_{t+1}$ is determined stochastically by the rule
\bel{2}
\sigma_{t+1} = \sigma_{t'} \quad \mbox{\rm with probability}\quad p\quad\mbox{and}\quad = -\sigma_{t'}\quad \mbox{\rm with } 1-p
\ee
At the first time step $t=1$ the process is initiated
as follows:\\ 
{\it (D3)} The elephant starting at $X_0$ moves to the right
with probability $q$ and to the left with probability $1-q$,
i.e.,
\bel{3}
\sigma_{1} = 1\quad \mbox{\rm with probabilty } q\quad\mbox{and}\quad = -1 \quad{\rm with }\quad 1-q.
\ee
It is obvious from the definition that
\begin{equation}
X_t = X_0 + \sum_{t'=1}^t \sigma_{t'}.
\label{l4}
\end{equation}
The question is to which extent the memory of the history influences the
distribution of the particle position. For $p<1/2$ the elephant behaves
metaphorically speaking
like a dedicated (but not very stringent) reformer:
At each step he is preferentially doing the opposite
of what he (randomly) remembers to have been decided in the past.
For $p>1/2$ the elephant is a more traditional type,
he preferentially sticks to his former decision.
Notice that three special cases of our model
are trivial: (i) In the borderline case $p=1/2$ the choice of $\sigma_{t+1}$
is $\pm 1$ with equal probability, no matter what the history was. Hence
one has the standard Markovian random walk which converges to Brownian motion
on large scales. In this case the initial parameter $q$
plays only a marginal role with no macroscopic significance.
(ii) In the limiting case $p=1$ the dynamics become essentially
deterministic. Given the first decision (which is random), the elephant moves
with probability 1 (ballistically) always one step in the same direction as
in the first move. Hence the first step is macroscopically relevant.
(iii) For $q=1/2$ the mean position of the elephant
is zero for all $p$. Nevertheless the distribution of $X_t$ depends
nontrivially on $p$.

To study the mean position $\exval{X_t}$ we first note that given the previous
history $\{\sigma_1,\dots,\sigma_t\}$ one has the conditional probability 
that the increment $\sigma_{t+1}$ takes the value $\sigma=\pm 1$
\begin{equation}
P[\sigma_{t+1} = \sigma |\sigma_1,\dots,\sigma_t] = \frac{1}{2t}
\sum_{k=1}^{t}
\left[1 + (2p-1)\sigma _k \sigma \right]\quad\rm{for}\quad t \geq 1
\label{5}
\end{equation}
This follows from the definitions {\it (D1), (D2)} of the process.
For $t=0$ we get in accordance with rule {\it (D3)}
\begin{equation}
P[\sigma_1 = \sigma ] = \frac{1}{2} \left[1 +(2q-1)\sigma \right]
\label{6}
\end{equation}
Thus for $t\geq 1$ the conditional mean increment is given by
\begin{equation}
\langle \sigma _{t+1}\mid \sigma _1 \dots \sigma _t \rangle
= \sum_{\sigma=\pm 1} 
\sigma P[\sigma _{t+1} = \sigma |\sigma_1,\dots,\sigma_t]
 = \frac{2p-1}{t} (X_t-X_0)
\label{7}
\end{equation}
These relations form the basis of the subsequent analysis of the process.
Below we shall frequently use the shifted parameters
\bel{8}
\alpha=2p-1, \quad \beta = 2q-1
\ee
which are in the range $[-1,1]$. Negative $\alpha$ corresponds
to the ``reformer'', positive $\alpha$ parametrizes the ``traditionalist''
elephant. The effectively memoryless Markovian case is $\alpha=0$. 
From (\ref{7}) the conventional mean value (obtained by summing over all
previous realizations of the process) is given by
\begin{equation}
\langle \sigma _{t+1} \rangle = \frac{\alpha}{t}\left(\exval{X_t}-X_0\right)
\label{9}
\end{equation}
and gives rise to the recursion for the mean displacement $\exval{x_t} =
\exval{X_t}-X_0$
\begin{equation}
\exval{x_{t+1}} = \left(1 + \frac{\alpha}{t}\right)\exval{x_t}
\mbox{ for } t\geq 1.
\label{10}
\end{equation}
For the first time step one has $\exval{x_{1}} = \exval{\sigma_{1}} =
2q-1=\beta$. The solution of (\ref{10}) is obtained by iteration
\begin{equation}
\langle x_t \rangle = \langle \sigma_1 \rangle
\frac{\Gamma (t+\alpha)}{\Gamma (\alpha+1)\Gamma (t)} \sim \frac{\beta}{\Gamma (\alpha+1)} t^{\alpha}\quad\mbox{for}\quad 
t\gg 1.
\label{11}
\end{equation}
For $\alpha<0$ (reformer) the mean displacement vanishes for large $t$
algebraically, the elephant stays on average essentially where it started.
For $\alpha>0$ (traditionalist) the mean displacement increases
indefinitely, albeit with decreasing velocity. The direction of the escape 
from the starting position is
determined by the first (random) decision. If the first move is positive,
the average direction of motion is to the right. Otherwise the elephant moves
on average to the
left. At the transition point $\alpha_c^{(1)}=0$ the mean displacement is
independent of time, as is known for the usual Markovian random walk. 
Recursion relations for higher order moments also follow straightforwardly
from (\ref{5}). They obey recursions of the form
\bel{13}
M_{t+1} = f_t + g_t\, M_t \mbox{ for } t\geq 1
\ee
where $M_t$ is some moment and $f_t,g_t$ are known functions, related
to lower moments. The general solution of (\ref{13}) is given by
\bel{14}
M_t = M_1\prod_{k=1}^{t-1}g_k + \sum_{n=1}^{t-1}
\left[f_n\prod_{k=n+1}^{t-1}g_k\right]
\ee
which is easily verified.
In particular for the second moment of the displacement one finds the 
recursion
\begin{equation}
\langle x^2_{t+1} \rangle = 1 +
\left(1 + \frac{2\alpha}{t}\right)\langle x^2_t \rangle
\label{15}
\end{equation}
Using (\ref{14}) it is solved by
\bel{16}
\exval{x^2_t} = \frac{t}{2\alpha-1}\left( \frac{\Gamma(t+2\alpha)}{\Gamma(t+1)
\Gamma(2\alpha)} - 1\right).
\ee
We first notice that the mean square displacement does not depend on the
initial decision parametrized by $q$ since $\exval{x^2_1}=1$ for
any $q$. Asymptotically one has
\bel{17}
\exval{x^2_t} = \frac{t}{3-4p}\quad p < 3/4;\quad \exval{x^2_t} = t \ln{t} \quad p = 3/4;\quad  
\exval{x^2_t} = \frac{t^{4p-2}}{(4p-3)\Gamma(4p-2)}\quad p > 3/4
\ee
Before discussing this result we remind the reader that the
displacement $x_t=X_t-X_0$ refers to the displacement from the initial
position, not to the displacement of the actual position from its mean.
Remarkably there is no qualitative change at $\alpha_c^{(1)}=0$ where
the transition to the escape regime occurs, yet
there are two distinct regimes inside the escape regime. \\
(1) For $\alpha < 1/2$ (corresponding to $p<3/4$) the mean square displacement 
increases asymptotically linearly in time. Hence the localized regime 
$\alpha<\alpha_c^{(1)}=0$ corresponds to a weak localization in the sense that 
the initial mean displacement vanishes for large $t$, but the variance 
increases diffusively with a diffusion coefficient $D = 1/(6-8p)$. In the 
range $0 \leq \alpha < 1/2$ (corresponding to $1/2 \leq p < 3/4$) the mean
displacement diverges (escape regime), but with an exponent $\alpha < 1/2$.
Therefore the mean square displacement  is still larger than the square of
the mean and the variance $\exval{x_t^2}-\exval{x_t}^2$ remains diffusive.\\
(2) For $\alpha > 1/2$ (corresponding to $p>3/4$) the mean square displacement
increases stronger than linearly $\sim t^{4p-2}$ and is of the
same order as the square of the mean, but with a different prefactor. 
Hence the variance becomes superdiffusive with an effective diffusion
coefficient depending both on time and on $q$.\\
(3) At the critical value $\alpha_c^{(2)}=1/2$ (corresponding to
$p=3/4$) the r.h.s. of (\ref{17}) reduces to $\sum_{n=1}^t t/n \sim t\ln{t}$.
The elephant is marginally superdiffusive.

The results of the previous section are sufficient for the
characterization of the large scale walk properties of the elephant only
if the increments $\sigma_n$ are independent random variables, i.e.,
for $\alpha=0$. In this case the central limit theorem guarantees convergence
of the distribution of $X_t$ to a Gaussian. In order to obtain information
about the distribution for $\alpha \neq 0$
we consider the complex-valued characteristic function $Q_t(k) = \exval{e^{ikx_t}}$. 
Using (\ref{5}) it obeys the equation
\begin{equation}
\langle Q_{t+1}(k) \rangle = \cos{k}\, Q_t(k)  +
\frac{\alpha}{t} \sin{k} \frac{d}{dk} Q_t(k).
\label{fp1}
\end{equation}
The Fourier transform $P_t(x)$ is the probability that the displacement at time $t$ takes
the value $x$. This is equal to the conditional probability $P(Y,t|X_0,0)$
that the position $X$ of the elephant at time $t$ equals $Y=X_0 + x$,
given that it started at $X_0$ at $t=0$. 
From (\ref{fp1}) we find a discrete evolution
equation formally analogous to the Fokker Planck equation (FPE) for usual 
random walks
\bea
P(Y,t+1|X_0,0) & = & \frac{1}{2} \left[1- \frac{\alpha}{t}(Y-X_0+1)\right]
P(Y+1,t|X_0,0) \nonumber \\
& & + \frac{1}{2} \left[1+ \frac{\alpha}{t}(Y-X_0-1)\right]
P(Y-1,t|X_0,0).
\label{fp2}
\eea
This equation (valid for $t\geq 1$) may be interpreted in terms
of a time-inhomogeneous random walk which does not memorize its full history,
but only its initial position at time $t=0$. It
describes a hopping process where in each step the walker at position
$Y$ hops to the right with probability $p_r = (1+\alpha(Y-X_0)/t)/2$
and to the left with probability $p_l = (1-\alpha(Y-X_0)/t)/2$
respectively. At first sight these stochastic dynamics look like a
time-inhomogeneous Markov chain where $X_0$ is some parameter.
However, we stress that in (\ref{fp2}) the quantity $X_0$ is not a parameter,
but the initial position of the elephant. The
hopping probabilities implicit in (\ref{fp2}) are not valid for
an elephant starting at a position different from $X_0$
or which starts at $X_0$ at a later time $t>0$.
The non-Markovian character of the dynamics is expressed in the fact that
the evolution equation (\ref{fp2}) is different for each initial position,
see \cite{Hang77} for a general discussion of similar non-Markovian
evolution equations. 
The qualitative features of the elephant which became apparent through the
study of its mean position are expressed in the hopping probabilities
$p_{r,\,l}$. For positive $\alpha$ the local bias
\be
b(x,t) = p_r - p_l = \frac{\alpha x}{t}
\ee
is positive for positive
displacement, hence the particle on average escapes. On the other hand,
for negative $\alpha$, the bias is opposite to the actual displacement,
reminiscent of some effective restoring force. This becomes very transparent
in the continuum limit (large displacement $x$ and time $t$). In terms
of $x,t$ (\ref{fp2})
takes the form
\begin{equation}
\frac{\partial P(x,t)}{\partial t} = \frac{1}{2}
\frac{\partial^2 }{\partial x^2} P(x,t) - \frac{\alpha}{t}
\frac{\partial }{\partial x} (x P(x,t)) \quad t >0\,
\label{fp3}
\end{equation}
of a FPE for a Brownian particle in a harmonic oscillator potential
with spring constant $k=\alpha/t$. The last relation is a special case of more 
general FPE with time-dependent coefficients that has been investigated in several papers 
\cite{lm,dh,mm}. Whereas the approach in those papers is phenomenologically we demonstrate 
in the frame of a microscopic model the origin of such FPE.   
From Eq.~(\ref{fp3}) one obtains recursion relations for the moments
of the distribution. Let us denote the even and the odd moments by
\be
a_n(t) = \exval{x^{2n}} \quad b_n(t) = \exval{x^{2n+1}}
  \quad \mbox{\rm{with} }  n = 0,1,2, \dots
\ee
Using Eq.~(\ref{fp3}) we obtain for the even moments
\begin{equation}
\frac{d}{dt}a_n(t) - \frac{2n\alpha }{t}a_n(t) = n(2n-1)a_{n-1}(t)
\label{mo1}
\end{equation}
and a similar equation for the odd moments. In particular, we have
\be
\frac{d}{dt}\langle x \rangle = \frac{\alpha}{t} \langle x \rangle
\ee
with the solution
\bel{mean}
\langle x(t) \rangle = \langle x(t_0) \rangle 
\left(\frac{t}{t_0}\right)^{\alpha} \equiv \bar{x}(t) \quad
\rm{with} \quad t\geq t_0 > 0
\ee
in agreement with (\ref{11}). Here $t_0$ is temporal cut-off scale,
reflecting the breakdown of the continuum approximation for $t \to 0$.
For second moment we find
\be
\exval{x^2} = \Delta(t) +  \bar{x}^2(t)
\ee
with $\bar{x}(t)$ given by (\ref{mean}) and
\bel{var}
\Delta(t) = \left(\frac{t}{t_0}\right)^{2\alpha}
\left[\exval{x^2(t_0)} - \bar{x}^2(t_0)\right] + 
\frac{t}{2\alpha-1} 
\left[ \left(\frac{t}{t_0}\right)^{2\alpha-1}-1 \right].
\ee
Since the initial distribution is assumed to be concentrated at $x_0$
the initial variance and so the first term in (\ref{var}) vanishes.
Thus we can read off the effective diffusion coefficient
\bel{diff}
D(t) = \frac{1}{4\alpha-2} 
\left[ \left(\frac{t}{t_0}\right)^{2\alpha-1}-1 \right]
\ee
of the elephant. 
In (\ref{mo1}) one recognizes the recursion relations for
the moments of a Gaussian distribution. Indeed,
one can straightforwardly verify that
\bel{dis}
P(x,t) = \frac{1}{\sqrt{4\pi tD(t)}} 
\exp{\left(-\frac{(x-\bar{x}(t))^2}{4tD(t)}\right)}
\ee
solves the evolution equation (\ref{fp1}) for the
initial condition $\delta(Y-X_0)=\delta(x- \bar{x}(t_0))$.
The centered even moments 
\be
M_{2n} = \exval{(x-\bar{x}(t))^{2n}} = (2n-1)!! (2tD(t))^n
\ee
which satisfy the recursion relation (\ref{mo1}) are given by the
usual expression for a Gaussian distribution.\\ 
The result can be generalized to the $d$-dimensional case with a separate
memory for each space direction. To this aim the rules {\it (D1)- (D3)} are generalized accordingly. 
Eq.~(\ref{l4}) is changed to an equation for a d-dimensional vector and correspondingly the conditional 
probability (\ref{fp2}) depends also on the d-dimensional position vector. In the continuous limit   
the evolution equation reads
\begin{equation}
\frac{\partial P(\vec x,t)}{\partial t} = \frac{1}{2}
\nabla ^2 P(\vec x,t) - \frac{\alpha}{t} \nabla (\vec x P(\vec x,t)).
\label{fp4}
\end{equation}
Based on that evolution equation we find also the equations for the even and
odd moments. In particular the
even moment $ a_n(t; d) = < (\vec x^2)^n>$ satisfies instead of 
Eq.~(\ref{mo1}) the equation
\begin{equation}
\frac{d}{dt}a_n(t;d) - \frac{2n\alpha}{t}a_n(t;d) = n [\,d + 2(n-1)]
a_{n-1}(t;d)
\label{mo6}
\end{equation}
A solution of this equation which yields the even centered moments is
\begin{equation}
a_n(t;d)= \frac{\Gamma (\frac{d}{2}+n)}{\Gamma (\frac{d}{2})}
(2tD(t))^n
\label{mo7}
\end{equation}
The odd moments $ b_n^{\beta }(t;d) = < (\vec x^2)^n x_{\beta }> $ obey the equation
\begin{equation}
\frac{d}{dt}b_n^{\beta }(t;d)) - \frac{(2n+1)\alpha}{t}b_n^{\beta }(t;d)) =
n(2n+d) b_{n-1}^{\beta }(t;d))
\label{mo8}
\end{equation}
for the odd moments of a $d$-dimensional Gaussian distribution.

Starting with a ``microscopic'' model of a random walk with unbounded
long-time memory (the ``elephant'') we have calculated the exact mean and 
variance respectively as well as the single-time probability distribution for 
the position of the elephant on large scales. Surprisingly
the memory effects incorporated in the probability distribution
at time $t$ amount to a time-inhomogeneous random walk where
only the initial position and starting time play a role.\\

\noindent {\bf Note added}: After submission of this paper, work on a similar model was published by S. Hod and U. 
Keshet, Phys. Rev. E {\bf 70}, 015104(R) (2004).

\begin{acknowledgments}
This work was supported by the DFG (SFB 418).
\end{acknowledgments}

\end{document}